\documentclass[twocolumn, superscriptaddress,showpacs,amsmath,amssymb,aps,pra,floatfix, 10pt]{revtex4-1}
\usepackage{dcolumn}
\usepackage{longtable}
\usepackage{mathtools}
\usepackage{graphicx}
\usepackage[american]{babel}
\usepackage[T1]{fontenc}
\usepackage{mathrsfs}
\usepackage{hyperref}
\usepackage{multirow}
\usepackage{bm, epsfig}
\usepackage{graphicx}
\usepackage{subeqnarray}
\usepackage{bbold}
\usepackage{upgreek}
\graphicspath{{Images/}}

\hypersetup{colorlinks=true, urlcolor=blue, citecolor=blue, filecolor=blue, linkcolor=blue}

\makeatletter
\newcommand*{\rom}[1]{\expandafter\@slowromancap\romannumeral #1@}
\makeatother

\begin{document}
\title{Finite Nuclear Size Effect to the Fine Structure of Heavy Muonic Atoms}
\author{Abu Saleh Musa Patoary}
\affiliation{Max Planck Institute for Nuclear Physics, Saupfercheckweg 1, 69117 Heidelberg, Germany}
\affiliation{Indian Institute of Technology, Kanpur, 208016 Kanpur, India}
\author{Natalia S. Oreshkina}
\affiliation{Max Planck Institute for Nuclear Physics, Saupfercheckweg 1, 69117 Heidelberg, Germany}

\date{\today} 

\begin{abstract}
The finite-nuclear size correction to the fine structure of muonic atoms are considered.
The procedure for the analytical calculation of the energies and wave functions has been derived in a homogeneously charged sphere nuclear 
  charge distribution approximation. 
The finite-nuclear size effect was calculated in a first few orders of the perturbation energy, with the accurate estimations of the 
  convergence.
Finally, we present energies of the low-lying electronic and muonic states with the finite-nuclear size correction, calculated analytically, 
  for ${}^{185}_{75}\text{Re}$ ion.
\end{abstract}


\maketitle

\section{Introduction}
One of the simplest and at the same time exotic atomic system is the highly charged hydrogen-like muonic atom, where a negative muon $\mu^-$ 
  is bound to an atomic nucleus.
Having the spin of $1/2$ as an electron does, a muon has a much larger mass, $m_\mu \approx 207 m_e$. 
Due to this fact, muonic orbits are closer to the nucleus than that of electron, making muonic systems a suitable probe to investigate and 
  extract nuclear parameters, i.e., nuclear size, charge distribution and moments~\cite{review of muonic atom}.

The study of muonic systems has been started a long time ago \cite{Jacobson1954, Acker1966}, mostly with a heavy system, where nuclear 
  effects are larger. 
For the light end of the periodic system, the most important and interesting is muonic hydrogen, where the prediction for the proton radius
	from the muonic spectroscopy disagree with the prediction based on the spectroscopy of the electronic hydrogen 
	(so-called proton-radius puzzle, see, e.g. \cite{nature_prp}).

In the present paper, we incorporate muonic systems into a procedure for analytical calculation of the finite nuclear size (FNS) effect 
  developed for electronic states \cite{outside nucleus}. 
The distribution of the nuclear charge is described by a the homogeneously charged sphere.
We show, that despite of the fact the FNS correction can not be considered as small, like it is in the case of an electronic system, 
  the analytical expressions derived from the Taylor expansion are still valid here.
The convergence of the series and the applicability limits for the procedure are discussed. 
The results for the first few orders of the FNS correction to the fine structure for the particular case of muonic ${}^{185}_{75}\text{Re}$ are presented. 
Our scheme can be successfully used for any other muonic atom. 

The paper is organized as follows. 
In Section \ref{electronic system} we demonstrate a procedure to estimate the FNS correction to the energies of fermionic states using analytically 
  derived wave functions. 
In Section \ref{muonic system} this procedure is modified to calculate FNS effect on ${}^{185}_{75}\text{Re}$ for all the states up 
  to principal quantum number $n=3.$ 
Relativistic units ($\hbar=c=1$) and Heaviside charge unit ($\alpha = e^2/4\pi$, where $\alpha$ is the fine structure constant, $e < 0$) are used in the paper.


\section{The solution of the Dirac equation for the extended nuclei} \label{electronic system}
One of the simplest models of nucleus is homogeneously charged sphere, with the corresponding charge density of nucleus
\begin{equation}
\rho(r) = \frac{3Ze}{4\pi r_0^3}\theta(r_0-r).
\end{equation} 
Here $Z$ is the nuclear charge
  and $r_0$ is effective radius of nucleus, connected with a root-mean-square (RMS) radius of the nucleus as \cite{justification of model}
\begin{equation}
r_0 = \sqrt{\frac{5}{3}\langle r^2\rangle}. \label{radius of sphere}
\end{equation} 
The interaction between electron and nucleus can be therefore described by the potential
\begin{equation}\label{eq:hom_sch_pot}
V(r)=
\begin{cases}
 -\frac{Z\alpha}{2r_0}(3-\frac{r^2}{r_0^2}), & \textrm{while } r \leq r_0 \textrm{ [Region \rom{1}}]; \\
 -\frac{Z\alpha}{r}, & \textrm{while } r>r_0 \textrm{ [Region \rom{2}}].
\end{cases}
\end{equation}

The energies $E$ and the wave functions $\psi(\textbf{r})$ of the bound fermion (electron or muon) are the eigenvalues and eigenfunctions, 
  respectively, of the stationary Dirac equation
\begin{equation} \label{eq:Dirac Equation}
\big[\pmb{\alpha} \cdot \pmb{p} + m_f \beta + V(r)\big]\psi(\textbf{r})=E\psi(\textbf{r}),
\end{equation}
where $m_f$ is the rest mass of fermion under consideration, $\pmb{\alpha},\beta$ are Dirac matrices, and $V(r)$ is the chosen central 
  nuclear potential, in our case, 
determined by Eq.~\eqref{eq:hom_sch_pot}.
For the arbitrary central potential, the radial and the angular dependence can be separated as
\begin{equation} \label{eq:wf}
\psi(\textbf{r})=
\binom{\frac{1}{r}G(r)\Omega_{\kappa m}(\theta,\phi)}{\frac{i}{r}F(r)\Omega_{-\kappa m}(\theta, \phi)},
\end{equation}
where $\kappa = (-1)^{j+l+\frac{1}{2}}(j+\frac{1}{2})$ is a relativistic angular quantum number, $j$ and $m$ are total angular momentum of 
  fermion and its projection, respectively, and $l$ is orbital angular momentum.
Angular part of the wave functions $\Omega_{\pm \kappa m}$ is the same for any central potential and well known~\cite{Relativistic QM}. 
Substituting expression~\eqref{eq:wf} into equation~\eqref{eq:Dirac Equation} and simplifying, one can obtain the system of radial equations
\begin{align}
\frac{dG}{dr}  +\frac{\kappa}{r}G(r) - [m_f-V(r)]F(r) =EF(r) \label{eq:G} \\
-\frac{dF}{dr} + \frac{\kappa}{r}F(r) + [m_f+V(r)]G(r) =EG(r)  \label{eq:F}
\end{align}

%


For the region \rom{1}, let us assume that the solution can be a power series of the form 
$$\binom{G(r)}{F(r)} = r^s\sum_{i=0}^\infty (a_i\pm b_i)r^i.$$ 
Taking the limit $r\rightarrow 0$ of equations \eqref{eq:G} and \eqref{eq:F}, one can find that $s = \pm|\kappa|.$ 
The solution corresponding to $s = -|\kappa|$ should be discarded to account for the regularity of the solution at origin. 
The remaining regular part is given by the formula
\begin{equation} \label{region 1}
\binom{G(r)}{F(r)} = N_1 r^{|\kappa|} \sum_{i=0}^\infty \big[a_i \pm (-1)^{i+1}\frac{\kappa}{|\kappa|}a_i\big]r^i,
\end{equation}
where $N_1$ is a free parameter and the coefficients are determined recurrently as
\begin{align}
a_i &= 0 \text{ for } i<0, \notag \\
a_0 &= 1, \notag \\
a_i &= \frac{a_{i-1}\big[E+\frac{3Z\alpha}{2r_0}-M(-1)^i\frac{\kappa}{|\kappa|}\big] - 
  \frac{Z\alpha}{2r_0^3}a_{i-3}}{\kappa+(-1)^{i+1}\frac{\kappa}{|\kappa|}(i+|\kappa|)}. \label{eq:coefficients_an}
\end{align}

   
To write a solution in region \rom{2}, we introduce a non-dimensional variable $\rho =  2r\sqrt{m_f^2-E^2} \equiv 2\lambda r.$ 
Representing $G,F$ by the linear combinations of two functions $\xi_1, \xi_2$ as
\begin{equation}
\binom{G(\rho)}{F(\rho)} = \sqrt{m_f\pm E}\rho^{-\frac{1}{2}}\big[\xi_1(\rho)\pm \xi_2(\rho)\big],
\end{equation}
one can rewrite Dirac equation to the system of the equations for $\xi_1,\xi_2$ as
\begin{align}
\rho\frac{d\xi_1}{d\rho} &= \xi_1 \bigg( \frac{\rho}{2}-q \bigg) - \xi_2 \bigg( \kappa+\frac{m_fZ\alpha}{\lambda} \bigg), 
  \label{variation of G} \\
\rho\frac{d\xi_2}{d\rho} &= \xi_1 \bigg( -\kappa+\frac{m_fZ\alpha}{\lambda} \bigg) + \xi_2 \bigg( q+1-\frac{\rho}{2} \bigg), 
  \label{variation of F}
\end{align}
where $q = {Z\alpha E}/{\lambda}-{1}/{2}$. 

Using the recursion relations for the Whittaker functions $W_{q,\gamma}(\rho)$ of the second kind \cite{mathematical function} one can show that the functions $\xi_1 = W_{q,\gamma}(\rho)$ and 
  {$\xi_2 = {W_{q+1,\gamma}(\rho)}/{(\kappa+\frac{m_fZ\alpha}{\lambda})}$}, where $\gamma = \sqrt{k^2-(Z\alpha)^2}$, satisfy equations \eqref{variation of G} 
  and \eqref{variation of F}.
Similarly, another solution for $\xi_1,\xi_2$ can be found using recursion relations between 
  $W_{-q,\gamma}(-\rho)$ and $W_{-q-1,\gamma}(-\rho)$. 
Two linearly independent solution sets allow one to describe any solution for $\xi_1, \xi_2$.
The condition of the regular behavior on infinity discards the divergent solution, leading to the following physical
solution outside of the nucleus

\begin{align}
\binom{G(\rho)}{F(\rho)} &= \frac{N_2}{\kappa + \frac{m_fZ\alpha}{\lambda}}\rho^{-\frac{1}{2}} \sqrt{m_f \pm E} \notag \\ 
&\times \left[ \bigg(\kappa + \frac{m_fZ\alpha}{\lambda} \bigg) W_{q,\gamma}(\rho) \pm W_{q+1,\gamma}(\rho) \right] \label{region 2},
\end{align}
where $N_2$ is again free parameter.
The expression \eqref{region 2} for the wave functions is in an agreement with \cite{outside nucleus}. 

	

So far, we presented the expressions for the wave function separately in two regions. 
As it corresponds to a physical state, the wave function should be normalized as
\begin{align*}
\int_0^{\infty}(G^2+F^2)dr=1,
\end{align*}
and it has be continuous at the boundary of two regions
\begin{align}
G(r_0-0) & = G(r_0+0),  \label{continuity of F}\\ 
F(r_0-0) & = F(r_0+0).  \label{continuity of G} 
\end{align}
$G$ and $F$ can be replaced by the explicit expression obtained in equation \eqref{region 1} and \eqref{region 2}, forming a homogeneous systems of linear equations on $N_1$ and $N_2$.
Non-trivial solution can be build only if the determinant of the system is equal to zero, which gives the following condition for the energy $E$
\begin{align} \label{eigenvalue}
&\frac{A_1 \big( \kappa+\frac{m_fZ\alpha}{\lambda} \big) W_{q,\gamma}(2\lambda r_0)+A_2W_{q+1,\gamma}(2\lambda r_0)}
{A_2 \big( \kappa+\frac{m_fZ\alpha}{\lambda} \big) W_{q,\gamma}(2\lambda r_0)+A_1 W_{q+1,\gamma}(2\lambda r_0)} \notag \\
& = \frac{\sum \limits_{i=0}^\infty a_ir_0^i}{\sum\limits_{i=0}^\infty (-1)^{i+1}\frac{\kappa}{|\kappa|} a_ir_0^i}, \\
&A_{1,2} = \sqrt{m_f+E} \pm \sqrt{m_f-E}. \notag  
\end{align}
By solving this equation, one can determine the energy of the electronic or muonic states in the field of the extended nuclei.


\subsection{Estimation of the energy for electronic systems}
Equation \eqref{eigenvalue} can not be solved analytically. 
To estimate the energy for the different states, let us assume that: \\
1. Energy shift ($\Delta E$) due to finite size of nucleus is small comparing to point-like-nucleus energy $E_0$, and 
  second and higher-order terms 
  can be neglected. \\
2. The infinite series in the right hand side (RHS) of equation \eqref{eigenvalue} is convergent. \\
Determining the finite nuclear size effect as  
\begin{align}
\delta E \equiv \frac{\Delta E}{E_0} = \frac{E-E_0}{E_0},
\end{align}
and applying the Taylor expansion around origin on the both sides of the \eqref{eigenvalue} up to the first  order, one can write
\begin{widetext}
\begin{align} \label{energy_1order}
\delta E = \cfrac{ \cfrac{\sum\limits_{i=0}^{\infty} a_i(0)r_0^i}{\sum\limits_{i=0}^{\infty}(-1)^{i+1}
  \cfrac{\kappa}{|\kappa|}a_i(0)r_0^i}  - \cfrac{F_1(0)}{F_2(0)}}  
    { \cfrac{F_1^{\prime}(0)F_2(0)-F_2^{\prime}(0)F_1(0)}{F_2(0)^2}  
	 - \cfrac{\sum\limits_{i,i'=0}^{\infty}a_i^{\prime}(0)a_{i'}(0)r_0^{i+i'}\cfrac{\kappa}{|\kappa|}\big[(-1)^{i}-(-1)^{i'}\big]}
	 {\bigg[\sum\limits_{i=0}^{\infty}(-1)^{i}\cfrac{\kappa}{|\kappa|}a_i(0)r_0^i\bigg]^2} } 
\end{align}
\end{widetext}

Here $F_1$ and $F_2$ are numerator and denominator, respectively, of the left hand side (LHS) of equation \eqref{eigenvalue}, expressed as a function of $\delta E$. 
Primed functions stand for the derivative with respect to $\delta E$, and the coefficients $a_n(\delta E)$ have been determined in equation \eqref{eq:coefficients_an}. 

As it was mentioned before, the formula for the FNS correction to the energy was derived in \cite{outside nucleus} for  electronic systems.
In this case, the FNS effects can be considered as a small correction, and therefore the terms of order $r_0^{2\gamma+1}$ has been safely neglected.
Terms of all orders in $r_0$ are included in equation \eqref{energy_1order}. 
To illustrate our statements, we calculated the energy shifts using formula \eqref{energy_1order} for several electronic ions and listed the results in Table~\ref{energy shift}.
Nuclear parameters have been taken from \cite{rmsradius}.

\begin{table}[h] 
\begin{center}
\begin{tabular}{|c|c|c|c|c|}
\hline
Ion & RMS (fm) & $E^{\rm B}_0$ (eV) & $\Delta E/E^{\rm B}_0$ & $E^{\rm B}$ (eV) \\
\hline
${}_{55}^{133}$Cs	& 4.8041	& 42986.8	& 0.000087(1)	& 42983.1 \\
${}_{75}^{185}$Re 	& 5.3596	& 83373.6	& 0.000425(3) 	& 83338.2(3) \\
${}_{81}^{205}$Tl 	& 5.4759	& 98880.6 	& 0.000678(1) 	& 98813.6(1) \\
${}_{82}^{208}$Pb 	& 5.5012	& 101641.1	& 0.000735(0)	& 101566.4(0) \\
${}_{92}^{238}$U 	& 5.8571	& 132361.3 	& 0.001723(2) 	& 132133.2(3) \\
\hline
\end{tabular}
\caption{
Dirac binding energy $E^{\rm B}_0$, the relative FNS correction to the energy $\Delta E/E^{\rm B}_0$, and the corrected binding energy $E^{\rm B}$ in some electronic ions with the uncertainty originated from RMS radius uncertainties. 
}
\label{energy shift}  
\end{center}
\end{table}

\begin{figure}[h]
\includegraphics[width=.99\columnwidth]{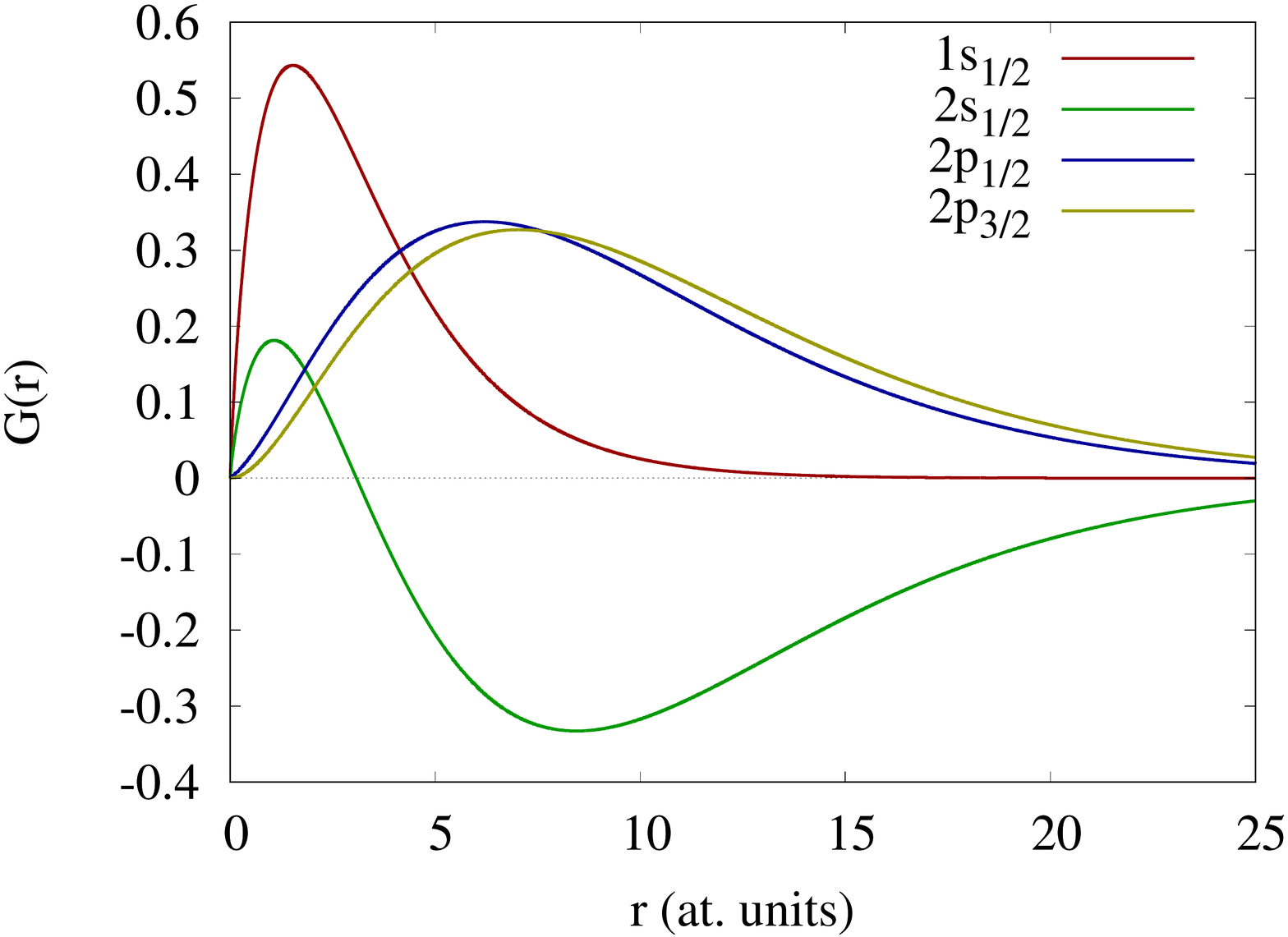}
\caption{(color online) The $G$ component of electronic radial wave function \eqref{eq:wf} calculated with the homogeneously charged nuclear model is plotted for four lowest lying states for hydrogen-like ${}^{185}_{75}\text{Re}$.}
\label{spherical nucleus: electronic wave function}
\end{figure}

After the calculation of the corrected energy one can use the equations  \eqref{region 1} and \eqref{region 2} to build the FNS corrected wave functions.
In Figure \ref{spherical nucleus: electronic wave function}, the upper radial component $G$ of the electronic 
wave functions \eqref{eq:wf} for four lowest lying states for ${}^{185}_{75}\text{Re}$ are shown. 
The series provided in equation  \eqref{region 1} converges very fast, and after the first few terms the next orders can be neglected.

\section{Nuclear Size Effect on Fine Structure of Muonic Atoms} \label{muonic system}

As an example of the muonic ion, let us consider the FNS correction to the fine structure of muonic ${}_{75}^{185}\text{Re}$ for the states up to $n=3$. 
Generally, all the formulas presented before for  electronic systems can be also used for the muonic systems by replacing electron's mass by muon's mass. 
However, since the FNS correction is not small anymore, the approximate formula \eqref{energy_1order} does not necessary provide a reasonable accuracy. 
Moreover, the analytical inclusion of the higher-order terms can be problematic in terms of convergence.
In this case, the Newton's method for the numerical solution of the equation was used.
The value $\delta E$ calculated by the formula \eqref{energy_1order} can be used normaly as a good initial approximation. 

\begin{figure}[h]
\includegraphics[width=.99\columnwidth]{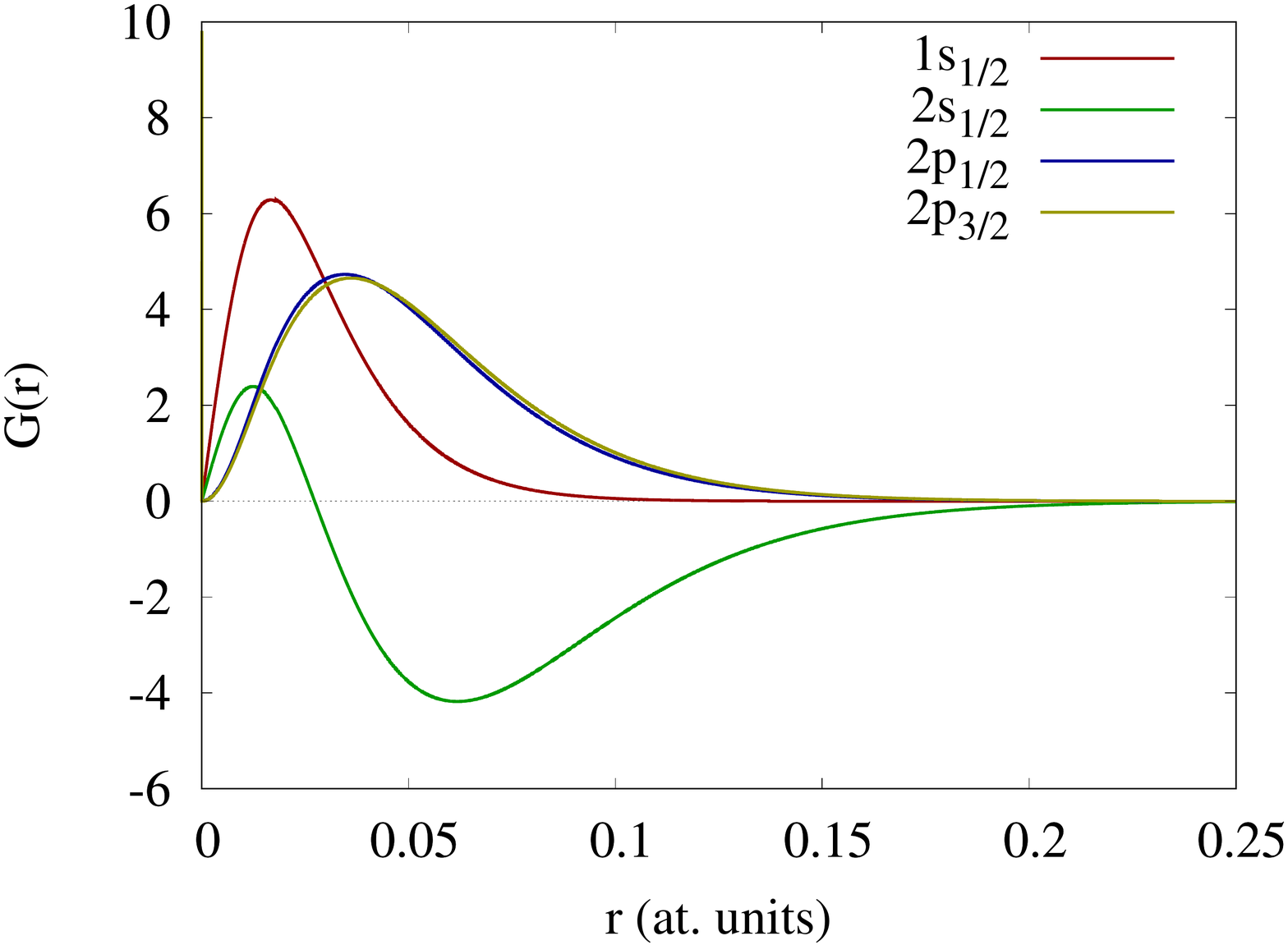}
\caption{The $G$ component of muonic radial wave function calculated \eqref{eq:wf} with the homogeneously charged nuclear model is plotted for four lowest lying states for hydrogen-like ${}^{185}_{75}\text{Re}$.}
\label{spherical nucleus: muonic wave function}
\end{figure}

In Table \ref{muonic Re-185}, we present FNS correction for ${}^{185}_{75}{\rm Re}$ up to the states with $n=3$. 
The first-order-contribution error bar appears due to the RMS uncertainty.
In Figure  \ref{spherical nucleus: muonic wave function}, the upper radial component $G$ of the muonic 
wave functions for four lowest lying states for ${}^{185}_{75}\text{Re}$ are shown. 
As it was expected, the FNS correction is more important for the $s$ states. 
Specifically for $1s_{{1}/{2}}$ state the FNS correction is almost 50\%, and definitely can not be considered as a small contribution. 
However, 
the high-accuracy results still can be delivered with the inclusions of the next-order terms.
This results could be considered as an excellent basis for further calculations of the fine-structure of highly-charged muonic ions.
\begin{table*}
\small
\begin{center}
\begin{tabular}{|c|r|l|l|l|l|l|}
\hline
State & $E^{\rm B}_0$ (MeV)& 
\multicolumn{4}{|c|}{
${\Delta E}/{E^{\rm B}_0}$} & $E^{\rm B}$ (MeV)\\
\hline
& & 1{st} order & 2{nd} order & 3{rd} order & 4{th} order & \\
\hline
$1s_{1/2}$ & 17.2286 & .4753(10) & -.0104(1) & -.0032 & -.0006 & 9.2845(190) \\
\hline
$2s_{1/2}$ & 4.3988 & .3025(6) & -.0006 &  & & 3.0708(26)  \\
$2p_{1/2}$ & 4.3988 & .0826(4) &   &  & & 4.0355(18) \\
$2p_{3/2}$ & 4.0328 & .0355(3) &  &  & & 3.8896(12) \\
\hline
$3s_{1/2}$ & 1.9129 & .2194(5) & -.0001 &  & & 1.4934(10) \\
$3p_{1/2}$ & 1.9129 & .0640(3) &  &  & & 1.7905(6) \\
$3p_{3/2}$ & 1.8039 & .0285(2) &  &  & & 1.7525(4) \\
$3d_{3/2}$ & 1.8039 & .0010 &  &  & & 1.8021 \\
$3d_{5/2}$ & 1.7730 & .0004 &  &  & & 1.7723 \\
\hline
\end{tabular}
\caption{Nuclear size correction factor in fine structure of muonic 
${}^{185}_{75}\text{Re}$ atom for the states up to $n=3$. 
$E^{\rm B}_0$ stands for Dirac value of the binding energy, ${\Delta E}/{E^{\rm B}_0}$ is energy shift due to finite nuclear size, $E^{\rm B}$ is the binding energy with FNS corrections. 
Errors in the first and second order correction is due to the uncertainty in RMS radius. 
Higher order contributions are listed in the cases when they are bigger than first order uncertainty.}
\label{muonic Re-185}
\end{center}
\end{table*} 

\section{Acknowledgements} 
The authors thank C.~H.~Keitel for the financial support, Z. Harman for the valuable advices and discussion, and N.~Michel for the comments about the manuscript.

\section{Author contribution statement}
N.S.O. conceived the presented idea. 
A.S.M.P. did the calculations. 
N.S.O. and A.S.M.P. wrote the main manuscript text.
All authors contributed to the discussion of the technical aspects and results, and preparation of the manuscript.


\begin{thebibliography}{50}

\bibitem{review of muonic atom}
E.~Borie, G.~A.~Rinker, Rev. Mod. Phys.  {\bf 54},  67 (1984)

\bibitem{Jacobson1954} 
B.~A.~Jacobsohn, Phys. Rev. {\bf 96},  1637 (1954)

\bibitem{Acker1966} 
H.~L.~Acker, Nucl. Phys. {\bf 87},  153 (1966)

\bibitem{nature_prp}
R.~Pohl {\it et al.} Nature {\bf 466},  213 (2010)

\bibitem{outside nucleus}
V.~M.~Shabaev, Opt. Spectr. {\bf 56},  244 (1984)

\bibitem{justification of model}
V.~M.~Shabaev, J. Phys. B {\bf 26},  1103 (1993)

\bibitem{Relativistic QM}
W. Greiner, Relativistic Quantum Mechanics, Springer (2000)

\bibitem{mathematical function}
M.~Abramowitz, I.~A.~Stegun, Handbook of Mathematical Functions and Formulas, Graphs and Mathematical Tables, National Bureau of Standarts (1964)


\bibitem{rmsradius}
I.~Angeli, K.~P.~Marinova, At. Data Nucl. Data Tabl. {\bf 99}, 69 (2013)

\end{thebibliography}
\end{document}